\documentclass[5p]{elsarticle}

\usepackage{bm} 
\usepackage{amsmath}

\usepackage{hyperref}

\journal{Physica B}

\begin{document}

\begin{frontmatter}

\date{14 October 2014}

\title{Tilted loop currents in cuprate superconductors}
\author{Victor M.~Yakovenko\corref{contact}}
\address{Department of Physics and Joint Quantum Institute, University of Maryland, 
 College Park, MD 20742-4111, USA}
\cortext[contact]{Tel: +1 301 405 6151;
 e-mail: yakovenk@physics.umd.edu \\
 URL: \url{http://physics.umd.edu/~yakovenk/}}
 
\begin{abstract}
The paper briefly surveys theoretical models for the polar Kerr effect (PKE) and time-reversal symmetry breaking in the pseudogap phase of cuprate superconductors.  By elimination, the most promising candidate is the tilted loop-current model, obtained from the Simon-Varma model by tilting one triangular loop up and another one down toward the apical oxygens.  The model is consistent with the PKE, spin-polarized neutron scattering, and optical anisotropy measurements.  Spontaneous currents in this model flow between the in-plane and apical oxygens in such a manner that each oxygen belongs to one current loop.  This loop-current pattern is similar to the spin order in the magnetoelectric antiferromagnet $\rm Cr_2O_3$, where the PKE is observed experimentally.  By analogy, it should be possible to train the PKE sign in the cuprates magnetoelectrically.  Several experiments are proposed to confirm the loop-current order: the magnetic-field-induced polarity, the nonlinear anomalous Hall effect, and the second-harmonic generation.
\end{abstract}


\end{frontmatter}


\section{Introduction}
\label{sec:Intro}

Experiments indicate that various symmetries are broken in the pseudogap phase of cuprate superconductors and several phase transitions separate it from the high-temperature metallic phase (see, e.g.,\ Fig.~1 in \cite{Taillefer}).  One of these broken symmetries is the time-reversal symmetry (TRS).  Spontaneous time-reversal symmetry breaking (TRSB) in cuprates was first proposed in the anyon superconductivity model \cite{Laughlin,Wen-Zee,Halperin}.  However, this model was discarded after negative experimental results for the Faraday effect in transmission \cite{Spielman1990} and the polar Kerr effect (PKE) in reflection \cite{Spielman1992} of light.  Eventually, the PKE was observed in the pseudogap phase with improved sensitivity of the specially-designed Sagnac interferometer \cite{Xia2008,Kapitulnik2009,He2011,Karapetyan2012,Karapetyan2014}.  The PKE in the cuprates appears well above the superconducting transition temperature and, apparently, is unrelated to superconductivity, unlike in the low-temperature superconductors $\rm Sr_2RuO_4$ \cite{Xia2006} and $\rm UPt_3$ \cite{Schemm2014}.

The PKE is routinely observed in ferromagnets, so its observation in the pseudogap phase of the cuprates was initially interpreted as evidence for a ferromagnetic-like order parameter, which breaks macroscopic TRS.  However, this interpretation faces difficulties, particularly in the view of the most recent measurements \cite{Karapetyan2014}.  The PKE is expected to have \emph{opposite} signs for light reflection from the opposite surfaces of a sample, because the ferromagnetic vector points into the sample for one surface and out of the sample for another surface.  However, the \emph{same} sign of the PKE was observed in \cite{Karapetyan2014}.  Moreover, it should be possible to control the PKE sign by going through the TRSB phase transition in the presence of an external magnetic field, as demonstrated in other materials \cite{Xia2006,Schemm2014}.  However, it was found that such ``training'' is extremely difficult \cite{Xia2008} or impossible \cite{Karapetyan2012,Karapetyan2014} to achieve in the cuprates.

These observations argue against macroscopic, ferro\-magnetic-like TRSB.  But the Sagnac interferometer is spe\-cifically designed to detect non-reciprocity in the normal reflection from a sample, which, according to Onsager's principle, is possible only if the TRS is broken.  These statements can be reconciled within antiferromagnetic-like models, where the TRS is broken microscopically, but not macroscopically.  In these models, local magnetic moments have opposite signs in two sublattices or within a unit cell, so the net ferromagnetic order parameter vanishes in the bulk, and the Faraday effect is absent in transmission of light.  Nevertheless, the PKE in reflection is generally non-zero, because it is primarily determined by magnetic moments in the first surface layer, as pointed out in \cite{Halperin}.  Magnetic moments of the opposite signs can be exposed at the opposite surfaces of a crystal, resulting in the PKE the \textit{same} sign upon reflections from these surfaces, in agreement with \cite{Karapetyan2014}.  Moreover, since the antiferromagnetic order parameter does not couple directly to an external magnetic field, training would be impossible or very difficult.

\begin{figure}
\centerline{ (a)\includegraphics[width=0.36\linewidth]{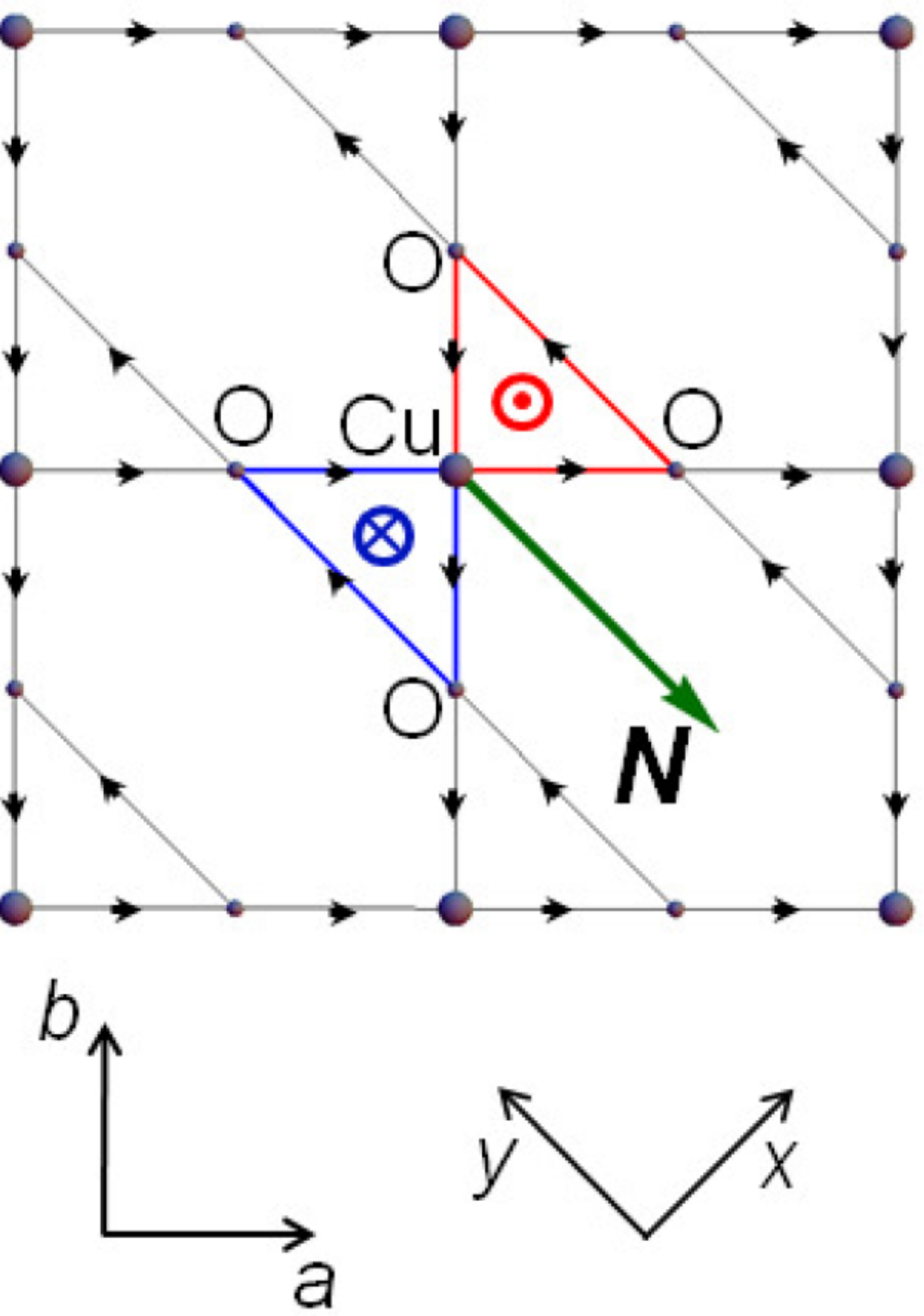} \hfill
(b)\includegraphics[width=0.5\linewidth]{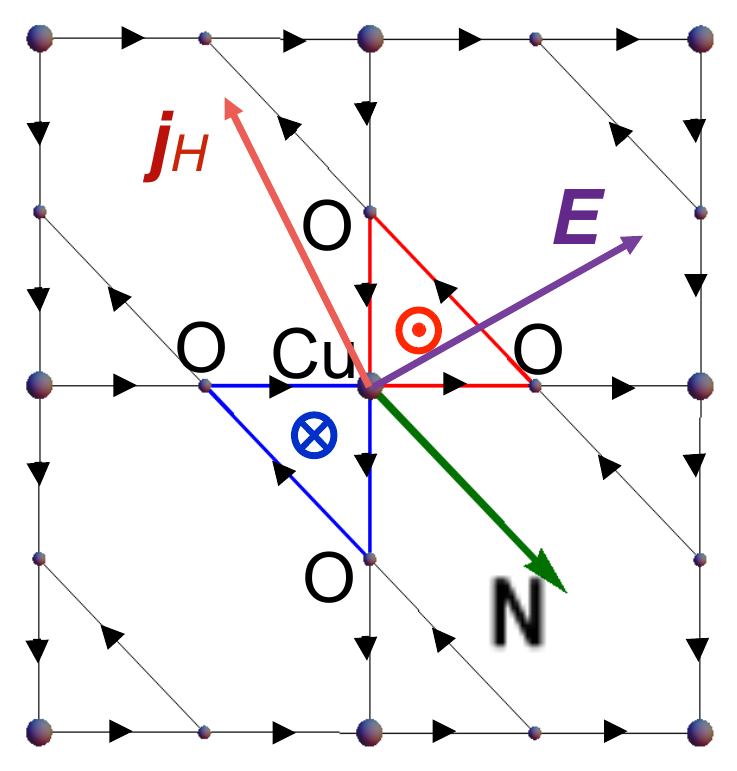} }
\caption{(a) Loop-current order in a CuO$_2$ plane \cite{Simon2002}.  Black arrows show directions of microscopic persistent currents between copper and oxygen atoms.  Green arrow shows the anapole moment $\bm N$. (b) The nonlinear anomalous Hall effect, where the in-plane Hall current $\bm j^H\propto\bm E\times[\bm E\times\bm N]$ is perpendicular to the applied in-plane electric field $\bm E$, but is proportional to the second power of $\bm E$.}
\label{fig:Varma}
\end{figure}

A symmetry analysis of whether the PKE is zero or non-zero in various antiferromagnetic models was presented by Orenstein \cite{Orenstein2011}.  Since antiferromagnetic ordering of atomic magnetic moments in the undoped phase is destroyed in the pseudogap phase, we focus on spontaneous orbital electric currents.  A much-discussed model of such loop currents was proposed by Simon and Varma \cite{Simon2002}.  This model is illustrated in Fig.~\ref{fig:Varma}(a), where the currents flow between the copper and oxygen atoms in the CuO$_2$ plane along triangular loops.  The orbital magnetic moments of the two loops have opposite directions along the $z$ ($\bm c$) axis perpendicular to the plane, so this is an orbital antiferromagnet with intra-unit-cell magnetic moments.  Substantial experimental support for this model was found in spin-polarized neutron scattering \cite{Fauque2006,Li2008,Li2011,Sidis2013,Mangin2014}.  These experiments also indicate that the net magnetic moment is zero, thus arguing against a ferromagnetic-like order parameter.  However, it was shown in \cite{Orenstein2011} by symmetry analysis that the model \cite{Simon2002} gives zero PKE.  Moreover, while the neutron scattering experiments do observe opposite magnetic moments inside the unit cell, these magnetic moments are not perpendicular to the CuO$_2$ plane, but are tilted by a substantial angle of the order of 45$^\circ$.

\begin{figure}
(a)\includegraphics[width=0.44\linewidth]{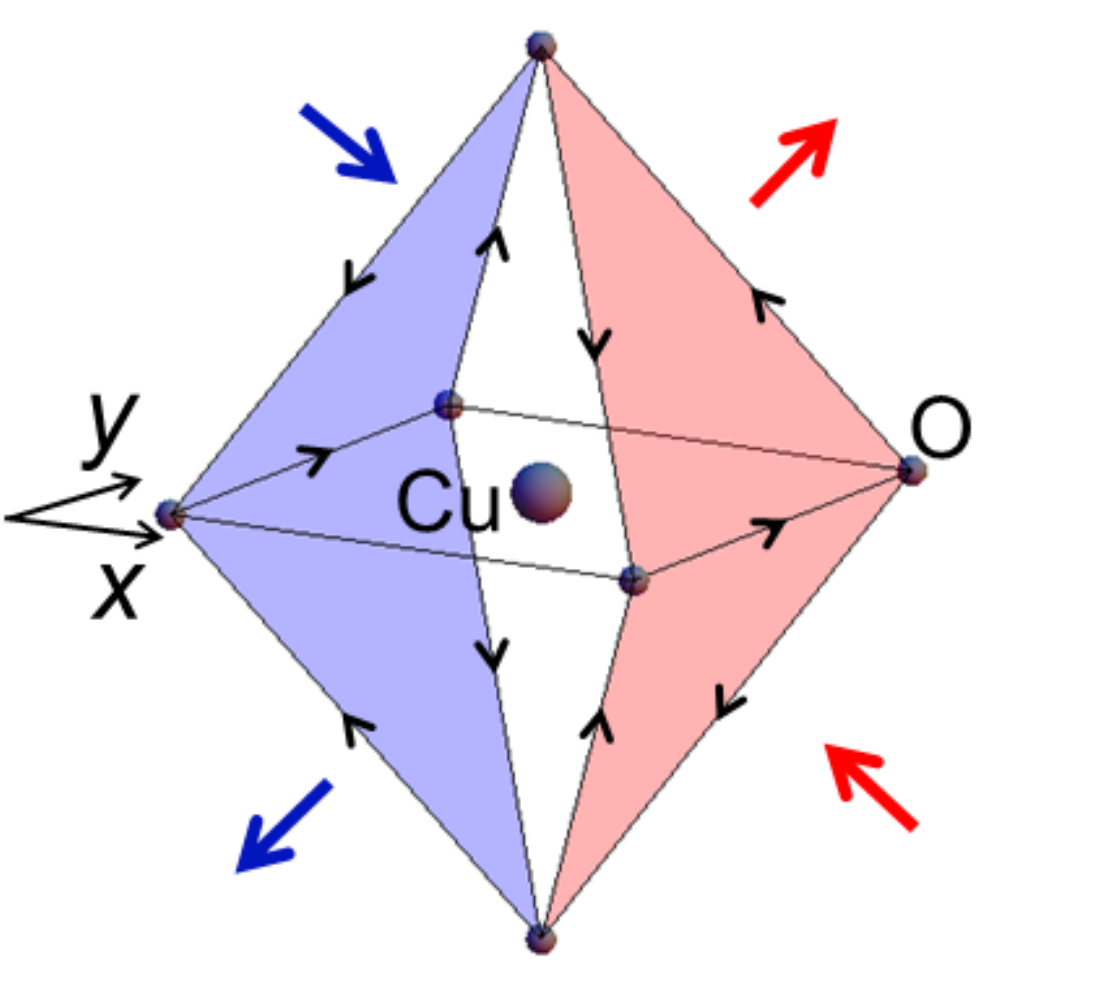} 
(b)\includegraphics[width=0.44\linewidth]{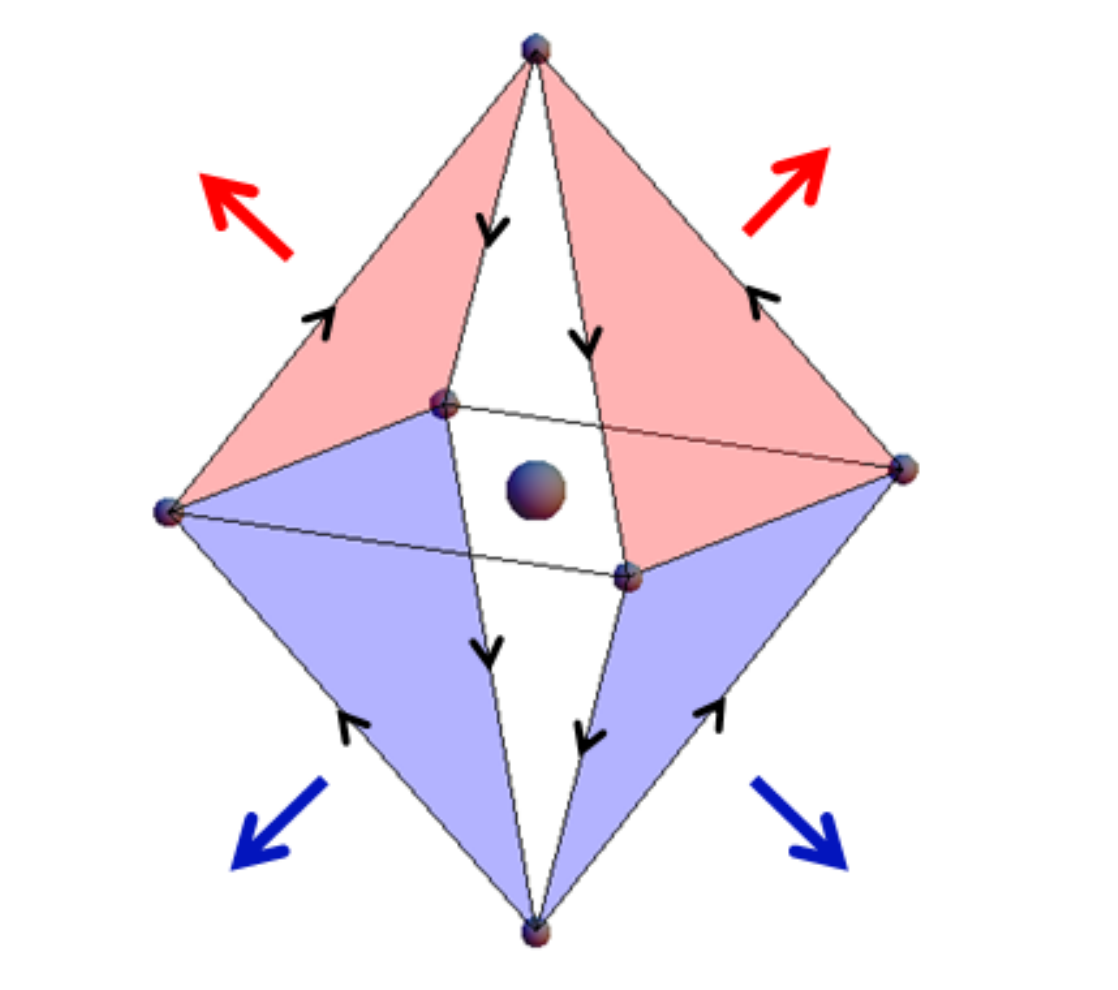} \\
(c)\includegraphics[width=0.44\linewidth]{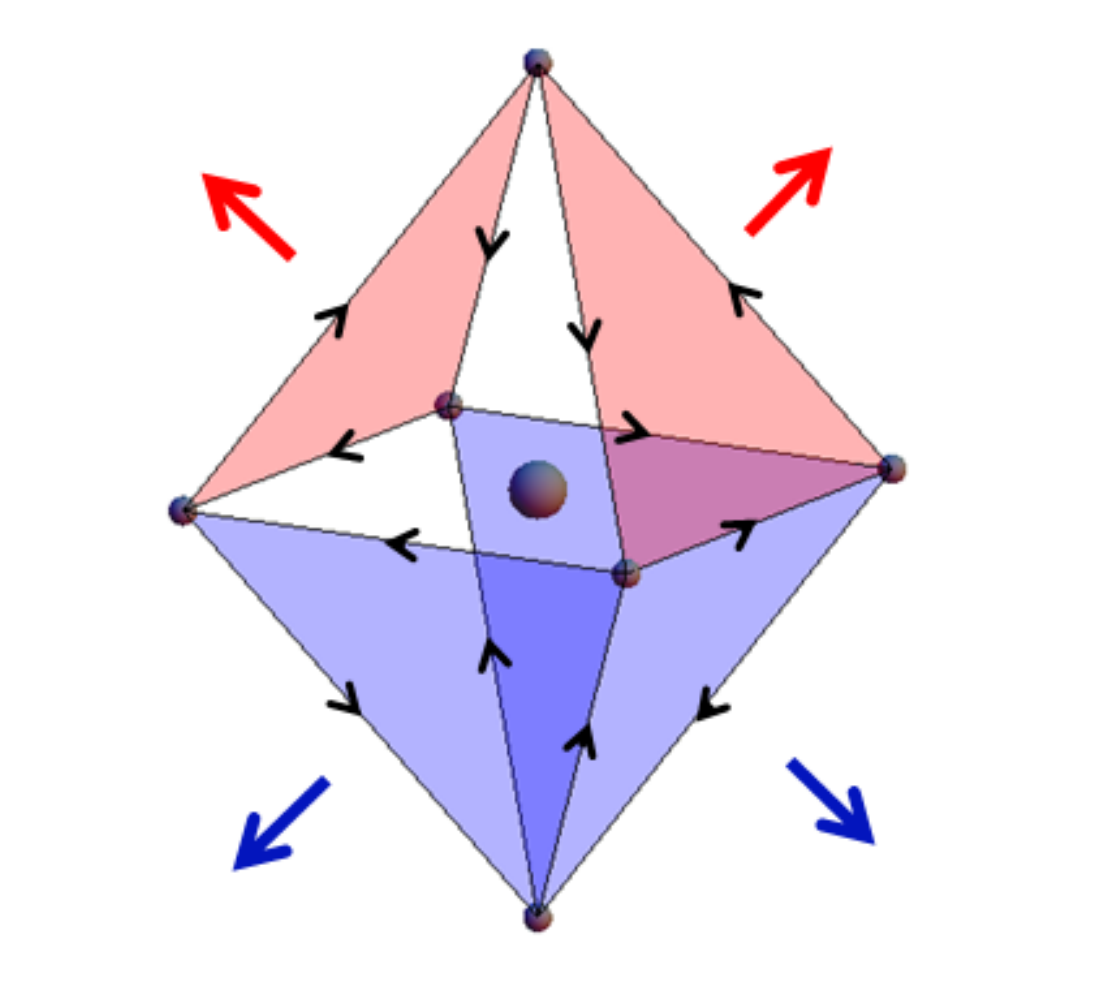} 
(d)\includegraphics[width=0.44\linewidth]{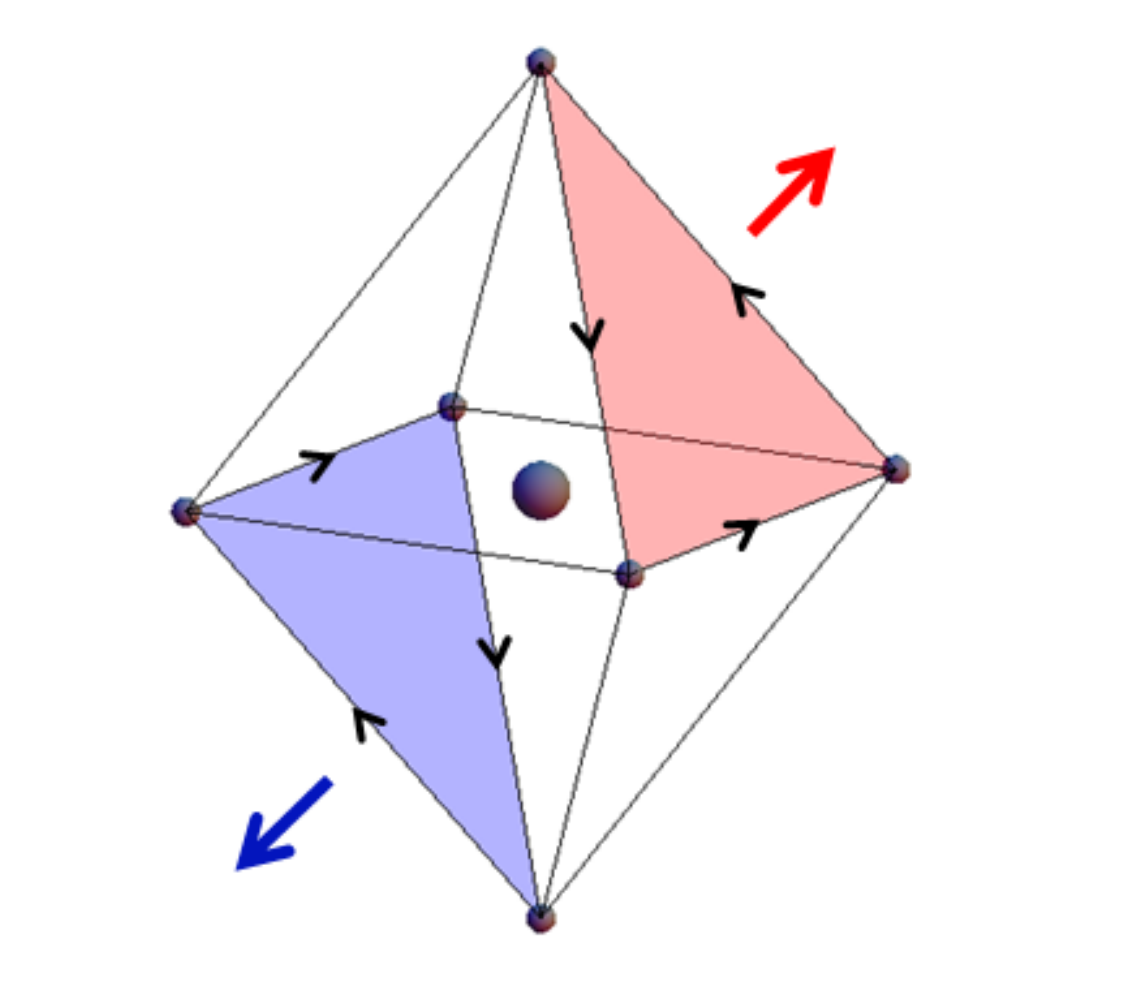}
\caption{Out-of-plane loop-currents models: (a) the model proposed in \cite{Weber2009}; (b) the model proposed in \cite{Orenstein2011} and shown in Fig.~2(a) there; (c) the model proposed in \cite{Orenstein2011}, but not shown there; (d) the titled loop-current (TLC) model proposed in Fig.~4.16(b) of \cite{Li-thesis}.  The red and blue arrows represent orbital magnetic moments with positive and negative $z$ components, respectively.} 
\label{fig:tilted}
\end{figure}

In order to explain the tilt of the magnetic moments, a modified model with out-of-plane loop currents was proposed in \cite{Weber2009} and also discussed in \cite{Li2008} in the context of $\rm HgBa_2CuO_{4+\delta}$.  This monolayer tetragonal material, where copper is surrounded by a octahedron of six oxygens, is conceptually the simplest and, thus, the most instructive to study \cite{Li2011}.  In the model of \cite{Weber2009}, shown in Fig.~\ref{fig:tilted}(a), the currents flow between the in-plane and the out-of-plane apical oxygen atoms located above and below the copper atom.  This structure can be obtained by duplicating the pattern shown in Fig.~\ref{fig:Varma}(a) and then buckling one copy up and another copy down.  In Fig.~\ref{fig:tilted}, the shading color, red or blue, of the triangular loops represents positive or negative projections of their magnetic moments onto the $z$ axis.  However, it was shown in \cite{Orenstein2011} by symmetry analysis that Fig.~\ref{fig:tilted}(a) also gives zero PKE.  It is also the case for an  alternative proposal \cite{He-Varma2012} to explain the tilt of magnetic moments by a quantum superposition of the states in Fig.~\ref{fig:Varma}(a) with four different orientations.  

In order to obtain a non-zero PKE, two other out-of-plane loop structures were proposed in \cite{Orenstein2011} and are shown in Figs.~\ref{fig:tilted}(b) and \ref{fig:tilted}(c).  Fig.~\ref{fig:tilted}(b) can be obtained from Fig.~\ref{fig:tilted}(a) by reversing the upper-left and lower-right loop currents, so that all magnetic moments point out of the octahedron formed by oxygen atoms.  Fig.~\ref{fig:tilted}(c) can be obtained from Fig.~\ref{fig:tilted}(b) by rotating the bottom loop currents by 90$^\circ$ around the vertical $z$ axis.  It was argued in \cite{Orenstein2011} that both Figs.~\ref{fig:tilted}(b) and \ref{fig:tilted}(c) give non-zero PKE.  However, they do not agree with the neutron scattering data \cite{Bourges}.  When Fig.~\ref{fig:tilted}(c) is viewed along the $z$ axis, i.e.\ projected onto the CuO$_2$ plane, the $z$ components of the magnetic moments form a $d$-wave pattern with alternating signs ${+- \atop -+}$ in the four quadrants of the $(\bm a,\bm b)$ plane \cite{Simon2002}, which does not agree with the neutron scattering.  When \ref{fig:tilted}(b) is viewed along the $z$ axis, the $z$ components of magnetic moments cancel out, and only in-plane components remain, in disagreement with the neutron scattering.

Yet another model, shown in Fig.~\ref{fig:tilted}(d), was discussed in \cite{Li-thesis}.  This structure can be obtained from Fig.~\ref{fig:Varma}(a) by tilting one triangular loop up and another triangular loop down.  Thus, I refer to this structure as the tilted loop-current (TLC) model.  Each oxygen atom belongs to only one current loop in Fig.~\ref{fig:tilted}(d), whereas the apical oxygens are shared between two current loops in the other panels of Fig.~\ref{fig:tilted}.  Fig.~\ref{fig:tilted}(d) can be thought of as a superposition of Figs.~\ref{fig:tilted}(a) and  \ref{fig:tilted}(b), where the upper-left and lower-right magnetic moments cancel out.  Since Figs.~\ref{fig:tilted}(a) and \ref{fig:tilted}(b) produce zero and non-zero PKE respectively, their superposition in Fig.~\ref{fig:tilted}(d) produces non-zero PKE.  Fig.~\ref{fig:tilted}(d) is reasonably consistent with neutron scattering \cite{Bourges,Li-thesis}.

Moreover, Fig.~\ref{fig:tilted}(d) agrees with the observation in \cite{Lubashevsky2014} that the in-plane principal optical axes are slightly rotated away from the crystallographic axes $\bm a$ and $\bm b$.  For a tetragonal crystal where the $\bm a$ and $\bm b$ axes are equivalent, the loop currents produce anisotropy between the $x$ and $y$ directions indicated in Fig.~\ref{fig:Varma}(a), so the principal optical axes are along the $\bm a+\bm b$ and $\bm a-\bm b$ directions.  This was shown in \cite{Varma2014} by calculating the renormalized dielectric susceptibility to the second order in the loop-current order parameter $\bm N$.  In addition, many cuprate materials have anisotropy between the $\bm a$ and $\bm b$ axes, either because of nematicity \cite{Taillefer,Kohsaka2007} or the crystal structure of YBCO.  A combination of the $\bm a$ vs.\ $\bm b$ and $\bm a+\bm b$ vs.\ $\bm a-\bm b$ anisotropies results in an intermediate orientation of principal axes away from the crystal axes.

In conclusion, among the loop-current models discussed above, only the TLC model shown in Fig.~\ref{fig:tilted}(d) is simultaneously consistent with the PKE, neutron, and optical anisotropy experiments.  However, all loop-current models are expected to produce non-zero magnetic fields at barium and oxygen sites \cite{Lederer2012},  whereas NMR measurements \cite{Strassle2011,Mounce2013,Mayaffre2014} find no such fields.  This discrepancy remains an open question.  

Sec.~\ref{sec:Multipole} of the paper briefly reviews multipolar expansion in electrodynamics of continuous media and its relation with magnetoelectric and non-reciprocal effects.  Sec.~\ref{sec:Cr2O3} compares the well-known magnetoelectric antiferromagnet $\rm Cr_2O_3$ with the loop-current models for cuprates.  Sec.~\ref{sec:SHG} proposes several experiments for independent confirmation of the loop-current order, namely the magnetic-field-induced polarity, the nonlinear anomalous Hall effect, and the second-harmonic generation.

\section{Multipole expansion for electromagnetic response}
\label{sec:Multipole}

Electrodynamics of continuous media can be developed by multipole expansion \cite{Raab-book}.  The first order in the multipole expansion includes the electric dipole density $\bm P$, and the second order includes the magnetic dipole $\bm M$ and electric quadrupole $Q_{\alpha\beta}$ densities \cite{Graham1997}:
  \begin{align} \label{PMQ}
  P_\alpha = & (\kappa_{\alpha\beta}-i\kappa'_{\alpha\beta})E_\beta 
  + \frac12(a_{\alpha\beta\gamma}-ia'_{\alpha\beta\gamma})\nabla_\gamma E_\beta \nonumber  \\
  & + (G_{\alpha\beta}-iG'_{\alpha\beta})B_\beta,   \\
  M_\alpha = & (G_{\beta\alpha}+iG'_{\beta\alpha})E_\beta , \quad
  Q_{\alpha\beta} = (a_{\gamma\alpha\beta}+ia'_{\gamma\alpha\beta}) E_\gamma,  \nonumber
  \end{align}
where $\bm E$ and $\bm B$ are the electric and magnetic fields with the time dependence $e^{-i\omega t}$.  The linear response tensors $\kappa$, $G$, and $a$ are the ensemble-averaged correlation functions between the microscopic electric dipole and electric dipole, magnetic dipole, and electric quadrupole operators, respectively \cite{Graham1992}.  Each tensor is separated into the terms that do ($\kappa'$, $a'$, $G$) or do not ($\kappa$, $a$, $G'$) change sign upon time reversal and either do ($a$, $a'$, $G$, $G'$) or do not ($\kappa$, $\kappa'$) change sign upon space inversion.  Generally, all these tensors are complex in the presence of dissipation.  Various terms in Eq.~(\ref{PMQ}) are permitted or forbidden for different (magnetic) symmetry classes of crystals.  The bound current density
  \begin{align} \label{J}
  J_\alpha^{(b)}=\dot P_\alpha -\frac12\nabla_\beta \dot Q_{\alpha\beta}
  + \epsilon_{\alpha\beta\gamma}\nabla_\beta M_\gamma
  \end{align}
is expressed in terms of Eq.~(\ref{PMQ}), and the bound charge density $\rho^{(b)}$ can be obtained from the continuity equation.  Electrodynamics of continuous media is obtained by substituting $\bm J^{(b)}$ and $\rho^{(b)}$ into Maxwell's equations.

At the electric-dipole level of the multipolar expansion, Eq.~(\ref{PMQ}) contains only the $\kappa$ tensor, which represents electric polarizability.  If TRS is not broken, only the tensor $\kappa_{\alpha\beta}=\kappa_{\beta\alpha}$ is permitted, and it is symmetric by Onsager's principle.  Principal axes of this tensor determine optical anisotropy of a material.  If the TRS is broken macroscopically, then the antisymmetric tensor $\kappa_{\alpha\beta}'=-\kappa_{\beta\alpha}'=-\sigma^H_{\alpha\beta}/\omega$ is permitted and is related to the Hall conductivity $\sigma^H_{\alpha\beta}$.  Various models with macroscopic TRSB, particularly in two dimensions (2D), where $\sigma^H_{xy}$ is determined by the Berry curvature and the Chern number, have been proposed \cite{Haldane1988,Yakovenko1990,Tewari2008,Sun-Fradkin2008,He-Moore2012,Aji-He2013,He-Lee2014}.  However, while $\sigma^H_{xy}$ does generate a non-zero PKE \cite[Sec.~101]{Landau}, the PKE signs are \textit{opposite} for the normally incident light from the top vs.\ from the bottom of the 2D plane, in contrast to the experiment \cite{Karapetyan2014}.  Thus, I discard the models with macroscopic TRSB and set $\kappa_{\alpha\beta}'=0=\sigma^H_{\alpha\beta}$.

The next term in Eq.~(\ref{PMQ}) with the coefficients $a$ and $a'$ represents spatial dispersion, i.e.,\ dependence of dielectric response on the wave vector $\bm k$ of light \cite[Ch.~12]{Landau}.  This term is permitted only if a material breaks inversion symmetry.  Let us first consider the case where the TRS is not broken, so only the terms $a$ and $G'$ are permitted in Eq.~(\ref{PMQ}).  Together, these terms are responsible for optical activity, i.e.,\ rotation of polarization of light on transmission through a non-TRSB chiral medium, such as sugar solution.  This scenario was proposed for cuprates, invoking the non-TRSB magnetoelectric tensor $G'$ \cite{Mineev2013} or the spatial-dispersion tensor $a$ \cite{Hosur2013,Orenstein2013,Pershoguba2013,Gorkov-1992}.  It was also pointed out that the electric quadrupole moment $Q_{\alpha\beta}$ should be taken into account on equal footing \cite{Norman2013}.  However, the authors later realized that these proposals cannot explain the PKE \cite{Mineev2014,Hosur2014,Pershoguba2014}, because it is forbidden by Onsager's principle in normal reflection \cite{Armitage2014,Fried2014}, if the TRS is not broken in any way (macroscopically or microscopically).  Thus, I set $a=G'=0$ below.

This leaves us with the terms $a'$ and $G$ in Eq.~(\ref{PMQ}), which are permitted if a material breaks the microscopic time-reversal and inversion symmetries, but preserves the combined symmetry.  This type of symmetry breaking is called magnetochiral \cite{Aji-He2013,He-Lee2014}.  It is characteristic \cite{Orenstein2011} for magnetoelectric antiferromagnets and loop-current models discussed in Sec.~\ref{sec:Intro}.  With the remaining terms $\kappa$, $a'$ and $G$ in Eq.~(\ref{PMQ}), the fields $\bm D$ and $\bm H$ can be expressed in terms of $\bm E$ and $\bm B$ by the following constituent relations \cite{Graham1997,deLange2014}
  \begin{align} \label{DH}
  D_\alpha = & A_{\alpha\beta}E_\beta + T_{\alpha\beta}B_\beta,   \\
  H_\alpha = & \mu_0^{-1} B_\alpha - T_{\beta\alpha}E_\beta,  \nonumber
  \end{align}
where
  \begin{align} 
   A_{\alpha\beta}= & \varepsilon_0\delta_{\alpha\beta}+\kappa_{\alpha\beta}
  -iS_{\alpha\beta\gamma}\nabla_\gamma,  
\label{A} \\
  S_{\alpha\beta\gamma}= & \frac13(a'_{\alpha\beta\gamma}+a'_{\beta\alpha\gamma}+a'_{\gamma\alpha\beta}),
\label{S} \\
  T_{\alpha\beta} = & G_{\alpha\beta} - \frac16 \omega \epsilon_{\beta\gamma\delta}a'_{\gamma\delta\alpha}.
\label{T}
  \end{align}
Generally, the microscopic definitions of the magnetic dipole and electric quadrupole correlators are not invariant with respect to an arbitrary shift of the origin of a coordinate system, but their combinations in Eqs.~(\ref{A}), (\ref{S}), and (\ref{T}) are invariant \cite{Graham1997,deLange2014}.  Using Eq.~(\ref{DH}) with the appropriate boundary conditions \cite{Graham1999,Graham2000}, the reflectivity matrix $R_{\alpha\beta}$ connecting the incident $E_\beta^{(i)}$ and reflected $E_\alpha^{(r)}=R_{\alpha\beta}E_\beta^{(i)}$ electric fields for normal incidence along the $z$ axis can be obtained \cite{Orenstein2011,Graham1999}.  The Sagnac interferometer detects violation of reciprocity due to TRSB by measuring the difference $R_{xy}-R_{yx}$ \cite{Fried2014}, which would otherwise be zero by Onsager's principle \cite{Halperin1992}.  It was shown in \cite{Orenstein2011} that
  \begin{align} \label{RT}
  \theta_K\propto(R_{xy}-R_{yx}) \propto (T_{xx}+T_{yy}),
  \end{align}
i.e.,\ the measured Kerr angle $\theta_K$ is proportional to the sum of the diagonal in-plane components of the magnetoelectric tensor $T$, whereas the tensor $S$ drops out.  It was shown in \cite{Orenstein2011} by symmetry analysis that $T_{xx}$ and $T_{yy}$ vanish for Figs.~\ref{fig:Varma}(a) and \ref{fig:tilted}(a), but are non-zero for Figs.~\ref{fig:tilted}(b) and \ref{fig:tilted}(c).  The claim about Fig.~\ref{fig:tilted}(c) is questionable, because this figure does have inversion symmetry.  Fig.~\ref{fig:tilted}(d), not discussed in \cite{Orenstein2011}, should have non-zero $T_{xx}$ and $T_{yy}$, because it is a superposition of Figs.~\ref{fig:tilted}(a) and \ref{fig:tilted}(b).  Thus, a non-zero PKE is permitted for Figs.~\ref{fig:tilted}(b) and \ref{fig:tilted}(d).

It was also found in \cite{Karapetyan2014} that $\theta_K$ changes linearly and antisymmetrically in response to applied uniaxial strains $u_{xx}$ and $u_{yy}$ in the $x$ and $y$ directions shown in Fig.~\ref{fig:Varma}(a), i.e., $\theta_K\propto(u_{xx}-u_{yy})$.  This result is consistent with the loop-current models in Figs.~\ref{fig:tilted}(b) and \ref{fig:tilted}(d).  Indeed, the strains $u_{xx}$ and $u_{yy}$ affect the projected loop-current order parameter in Fig.~\ref{fig:Varma}(a) linearly, but with different coefficients.  Thus, $\theta_K$ changes linearly in response to $u_{xx}-u_{yy}$.  In contrast, the strains $u_{aa}$ and $u_{bb}$ affect the projected loop-current order parameter  in Fig.~\ref{fig:Varma}(a) in the same manner (for a tetragonal material), so $\theta_K$ does not respond to $u_{aa}-u_{bb}$ in the first order.

\section{Comparison between $\rm Cr_2O_3$ and the cuprates}
\label{sec:Cr2O3}

From the above consideration, it is clear that the magnetoelectric tensor $T$ is crucial for the PKE.  The classic, much-studied magnetoelectric antiferromagnet is $\rm Cr_2O_3$, so it is instructive to discuss its properties in comparison with the cuprates.  The magnetoelectric effect in this material was predicted theoretically in \cite{Dzyaloshinskii1960} and subsequently observed experimentally in \cite{Astrov1960,Folen1961,Rado1961}.  The PKE in $\rm Cr_2O_3$ was predicted in \cite{Hornreich1968} and observed in \cite{Krichevtsov1993,Krichevtsov1996}.  The signal observed in \cite{Krichevtsov1993} was several orders of magnitude greater than theoretically estimated in \cite{Hornreich1968}.  Apparently, it is the theoretical underestimate that discouraged experimentalists from trying to observe the PKE earlier, until the interest was stimulated by  cuprate superconductors \cite{Spielman1992}.  Moreover. it was found that the PKE is enhanced at frequencies near certain optical transitions, and both real and imaginary parts of the signal were measured \cite{Krichevtsov1996}.

\begin{figure}
\centerline{\hfill (a)\includegraphics[height=0.6\linewidth]{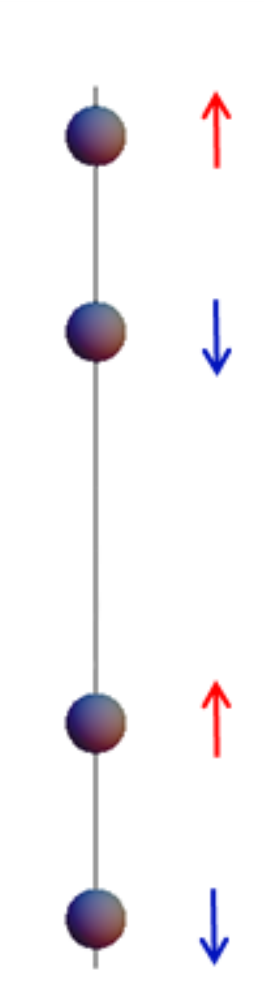} 
\hfill (b)\includegraphics[height=0.6\linewidth]{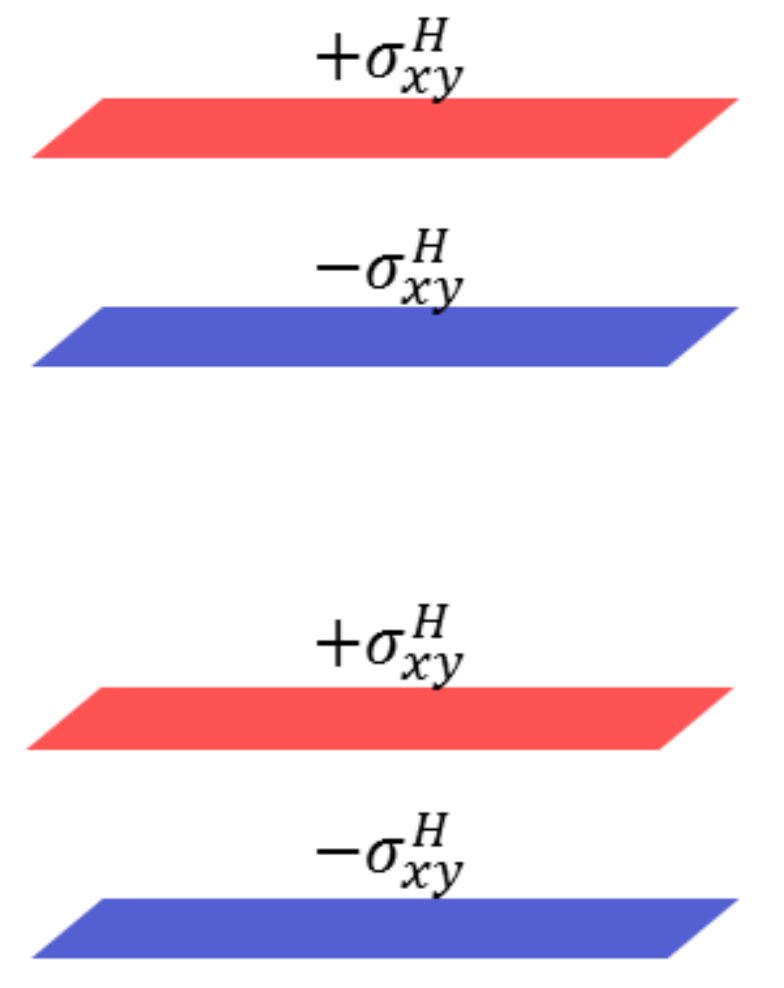} \hfill}
\caption{(a) Stylized structure of Cr$_2$O$_3$. (b) An antiferromagnetic anyon model discussed in \cite{Dzyaloshinskii1991,Canright-PRB1992}.}
\label{fig:Cr2O3}
\end{figure}

The structure of $\rm Cr_2O_3$ is illustrated in Fig.~\ref{fig:Cr2O3}(a) in a highly stylized manner.  The chromium atoms are grouped in pairs, where the two atoms with the opposite magnetic moments along the $z$ axis are displaced relative to each other along the $z$ axis.  Dzyaloshinskii \cite{Dzyaloshinskii1991} pointed out that the symmetry of $\rm Cr_2O_3$ is similar to the antiferromagneti\-cally-ordered anyon model illustrated in Fig.~\ref{fig:Cr2O3}(b), where the two $\rm CuO_2$ planes within a bilayer have opposite signs of the spontaneous Hall conductivity $\pm\sigma^H_{xy}$.  Heuristically, one may think that the staggered magnetic moments in Fig.~\ref{fig:Cr2O3}(a) produce alternating signs of the Hall conductivity in Fig.~\ref{fig:Cr2O3}(b).  The PKE for the model in Fig.~\ref{fig:Cr2O3}(b) was calculated in \cite{Dzyaloshinskii1991} and further investigated in \cite{Canright-PRL1992,Canright-PRB1992,Dzyaloshinskii1995}.

It was shown in \cite{Dzyaloshinskii1960} that $\rm Cr_2O_3$ has the following non-zero components of the magnetoelectric tensor $G_{xx}=G_{yy}\neq G_{zz}$.  Heuristically, the origin of these tensor components can be illustrated as follows.  Suppose the magnetic moments in Fig.~\ref{fig:Cr2O3}(a) are not fully saturated.  Applying an electric field $E_z$ would transfer some electron density from blue to red atoms, thus breaking the balance between their magnetic moments and inducing magnetization $M_z=G_{zz}E_z$.  The components $G_{xx}$ and $G_{yy}$ are easier to illustrate in Fig.~\ref{fig:Cr2O3}(b), which is equivalent by symmetry to Fig.~\ref{fig:Cr2O3}(a).  An in-plane electric field $E_y$ induces the transverse currents $J_x=\pm\sigma^H_{xy}E_y$ in the planes.  When these currents are closed within bilayers at infinity, they produce magnetization $M_y=G_{yy}E_y$.

The TLC pattern in Fig.~\ref{fig:tilted}(d) is qualitatively similar to Fig.~\ref{fig:Cr2O3}(a), with the orbital magnetic moments instead of spins.  The common building block in these figures is a pair of opposite magnetic moments with non-zero projections onto the vector of their relative displacement.  This motif is doubled in Fig.~\ref{fig:tilted}(b), but canceled in Fig.~\ref{fig:tilted}(a), where the magnetic moments point outwards in one pair, but inwards in another pair.  It was originally thought that the pattern in Fig.~\ref{fig:Cr2O3}(b) applies only to bilayer cuprates \cite{Dzyaloshinskii1991}, but, taking into account the top and bottom apical oxygens, it also applies to monolayer compounds as shown in Fig.~\ref{fig:tilted}.  To summarize, in the presence of the TLC pattern in Fig.~\ref{fig:tilted}(d), the cuprates become magnetoelectric orbital antiferromagnets with the properties similar to $\rm Cr_2O_3$, where the PKE was observed experimentally.

Now let us discuss the issue of domains.  In Fig.~\ref{fig:tilted}, different states can be obtained by space rotations and by reversing the currents, thus domains of different states are possible.  In experiments \cite{Krichevtsov1993,Krichevtsov1996}, $\rm Cr_2O_3$ was successfully trained magnetoelectrically to produce a single domain.  External parallel magnetic and electric fields were applied to $\rm Cr_2O_3$ when passing through the antiferromagnetic phase transition and then switched off at low temperatures when measuring the PKE.  It would be interesting to apply this magnetoelectric training to the cuprates.

Generally, for any spontaneous symmetry breaking, there are, at least, two different states of equal energies connected to each other by the symmetry operation that is broken.  Thus, it is possible to have domains of these different states.  Nevertheless, the possibility of domains does not preclude experimentalists from successfully observing phase transitions with spontaneous symmetry breaking in macroscopic measurements, as it has been demonstrated in great many cases.  The often-quoted examples of ferromagnets or ferroelectrics are special.  A monodomain structure in these materials would result in macroscopically large energy of magnetic or electric fields due to their long-range character.  Thus, domain proliferation is inevitable for ferromagnets and ferroelectrics.  However, this argument does not apply to antiferromagnetic and other order parameters that do not produce long-range fields.  There is no fundamental reason for proliferation of domains in these cases, because domain walls have positive energy.

\section{Magnetic-field-induced polarity, nonlinear Hall effect, 
and second-harmonic generation}
\label{sec:SHG}

In the original Simon-Varma model shown in Fig.~\ref{fig:Varma}(a), the out-of-plane magnetic moments have zero projections onto the in-plane vector connecting them.  Thus, the diagonal magnetoelectric coefficients vanish $G_{xx}=G_{yy}=G_{zz}=0$ \cite{Orenstein2011} and the PKE is zero.  However, it is well known \cite{Shekhter-2009} that this model has non-zero off-diagonal magnetoelectric coefficients $G_{xz}\neq G_{zx}$, where the $x$ and $y$ axes are shown in Fig.~\ref{fig:Varma}(a).  Although $G_{xz}$ and $G_{zx}$ do not contribute to the PKE, they have other experimental consequences discussed in this Section.  For simplicity, only $G_{xz}$ is considered below.  Let us introduce the anapole vector $\bm N$ as in \cite{Shekhter-2009}, shown by the green arrow in Fig.~\ref{fig:Varma}(a),
\begin{equation}
	\bm N = \int d^2r\, [\bm m(\bm r)\times\bm r]
	= \frac{1}{2c} \int d^2r\, r^2Ê\bm j(\bm r),
	\label{eqL}
\end{equation}
where $\bm m(\bm r)$ and $\bm j(\bm r)$ are the microscopic densities of the magnetic moment and electric current, and the integral is taken over the unit cell.  Then, Eq.~(\ref{PMQ}) gives
  \begin{align} \label{ME}
  \bm P = G_{xz}\,[\bm B_z\times\bm n],  \quad  \bm M_z = G_{xz}\,[\bm n\times\bm E]_z,
  \end{align}
where $\bm P$ and $\bm E$ are in-plane vectors, and $\bm n=\bm N/N$.

Nematicity, i.e.,\ spontaneous anisotropy between the $\bm a$ and $\bm b$ axes, is often observed in STM measurements in cuprates \cite{Kohsaka2007}.  It means that the two oxygens on the vertical axis in Fig.~\ref{fig:Varma}(a) are inequivalent to the two oxygens on the horizontal axis.  Yet, within each pair, the oxygens remain equivalent, and extensive analysis of STM data failed to find any evidence for inversion symmetry breaking \cite{Seamus}.  However, if an out-of-plane magnetic field $B_z$ is applied, it would induce an in-plane dipole moment $P_x$ according to the first Eq.~(\ref{ME}).  This would make the top oxygen inequivalent to the bottom one, and the right oxygen inequivalent to the left one.  So, the theory predicts that a strong out-of-plane magnetic field should induce in-plane polarity, observable by STM.

Now suppose that an in-plane electric field is applied.  According to the second Eq.~(\ref{ME}), it induces the out-of-plane magnetic moment $M_z$, which, from symmetry standpoint, is equivalent to an effective magnetic field $B_z^{\rm eff}$.  The electric and effective magnetic field together produce the in-plane Hall current $\bm j^H$, as shown in Fig.~\ref{fig:Varma}(b),
  \begin{align} \label{Hall}
  \bm j^H \propto (\bm E\times\bm B^{\rm eff}) 
  \propto (\bm E\times[\bm E\times\bm N]),
  \end{align}
where $\bm j^H$, $\bm E$, and $\bm N$ are in-plane vectors.  Eq.~(\ref{Hall}) represents the nonlinear anomalous Hall effect: the current $\bm j^H$ is perpendicular to the applied electric field $\bm E$, but it is proportional to the second power of $\bm E$, and no external magnetic field is applied.  It was discussed in terms of the Berry curvature in \cite{Moore2010,Gao2014}, with applications to GaAs quantum wells \cite{Moore2010}.  Our proposal is that the nonlinear anomalous Hall effect should be observable in cuprates in the presence of the loop-current order.  

If an ac electric field $\bm E(\omega)$ at the frequency $\omega$ is applied, then Eq.~(\ref{Hall}) produces two different effects.  One effect is the second-harmonic generation, where the current $\bm j^H(2\omega)$ is induced at the frequency $2\omega$, obtained by combining $\bm E(\omega)$ and $\bm E(\omega)$.  It is well-known that the second-harmonic generation is permitted only when the inversion symmetry is broken \cite[Ch.~13]{Landau}.  In the loop-current models, the vector $\bm N$ changes sign upon inversion, thus Eq.~(\ref{Hall}) is permitted.  However, because the vector $\bm N$ also changes sign upon time reversal, no evidence for inversion breaking is observed in STM in the absence of an external magnetic field.  Another effect following from Eq.~(\ref{Hall}) is rectification of the dc current $\bm j^H$ produced by combining $\bm E(\omega)$ and $\bm E(-\omega)$, known as the photogalvanic effect \cite{Gorkov-1992,Moore2010,Hosur2011}.  If a short pulse of high-frequency radiation is applied, the current $\bm j^H$ would change in time as the envelope of this pulse.  

Although the above discussion was presented for the in-plane loop currents in Fig.~\ref{fig:Varma}(a), the same conclusions apply to the TLC model shown in Fig.~\ref{fig:tilted}(d).   Moreover, the TLC model has additional features due to diagonal magnetoelectric coefficients.  Using $G_{xx}$, an in-plane electric field $E_x$ induces an effective in-plane magnetic field 
$B_x^{\rm eff}\propto G_{xx}E_x$.  Then, an in-plane electric field $E_y$ induces the Hall current along the z axis:
  \begin{align} \label{Hall-z}
  j_z^H \propto E_y B_x^{\rm eff} 
  \propto G_{xx}E_xE_y \propto E^2\cos(2\phi),
  \end{align}
where $\phi$ is the in-plane azimuthal angle measured from the $\bm a$ axis in Fig.~\ref{fig:Varma}(a).  Eq.~(\ref{Hall-z}) shows that the dc out-of-plane anomalous Hall current is maximal and has opposite signs for the in-plane ac electric field polarized along the $\bm a$ or $\bm b$ axes.  A similar effect obtained from $G_{yy}$ does not cancel out, because $G_{xx}\neq G_{yy}$ for the TLC model.

Experimental observation of these nonlinear optical effects in the cuprates would strongly support the loop current models discussed in this paper.  It is worth mentioning that the second-harmonic generation was experimentally observed in $\rm Cr_2O_3$ \cite{Fiebig-1994}.  Opposite signs of the effect were detected in different antiferromagnetic domains by scanning with a laser spot, and it was found that the same domain can penetrate through the bulk and terminate at the opposite surfaces of the crystal \cite{Fiebig-1994}.  A review of the second-harmonic generation in magnetically ordered crystals is given in \cite{Fiebig-2005}.

\section{Conclusions}

This brief review critically examines various models proposed for theoretical explanation of the PKE and TRSB observed in the pseudogap phase of cuprates.  By elimination, the paper concludes that the most promising candidate is the TLC model shown in Fig.~\ref{fig:tilted}(d).   This model is simultaneously consistent with the PKE, spin-polarized neutron scattering, and optical anisotropy measurements.  The pattern of orbital loop currents in the TLC model in Fig.~\ref{fig:tilted}(d) is similar to the spin order in the magnetoelectric antiferromagnet $\rm Cr_2O_3$ in Fig.~\ref{fig:Cr2O3}(a), where the PKE has been observed experimentally.  By analogy with the experiments in $\rm Cr_2O_3$, it is proposed that the PKE sign in the cuprates can be trained by applying magnetic and electric fields simultaneously, rather than only magnetic field.  Several experiments are proposed for independent confirmation of the loop-current order in the cuprates: the magnetic-field-induced polarity, the nonlinear anomalous Hall effect, and the second-harmonic generation.  The latter effect has been experimentally observed in $\rm Cr_2O_3$.

The paper focuses only on phenological identification of a model consistent with the experiments, but leaves a possible microscopic justification of such a model to future studies.  The variational Monte Carlo study \cite{Weber2009} has shown that the out-of-plane loop-current order in Fig.~\ref{fig:tilted}(a) may be a viable contender for the ground state of the system.  However, the mirror reflection symmetry $z\to-z$ was imposed on the trial wave functions in \cite{Weber2009}.  If this condition is relaxed and lower-symmetry states are permitted, the variational Monte Carlo method can be applied the TLC state in Fig.~\ref{fig:tilted}(d).  The loop-current pattern of \cite{Weber2009} was studied by exact diagonalization on limited-size clusters in \cite{Kung2014} and found to be not viable, but the TLC pattern of Fig.~\ref{fig:tilted}(d) was not investigated.
Many other open questions remain in microscopic theory, including a possibility of using oxygen orbital moments instead of circulating loop currents \cite{Moskvin2012}.  This review focuses only on breaking of discrete symmetries, such as TRS, inversion, and rotation, while preserving translational symmetry and not changing the unit cell.  However, the charge-density wave is known to appears in the cuprates, arguably at a lower temperature that the pseudogap transition (see Fig.~1 in \cite{Taillefer}), breaking translational symmetry, folding the Brillouin zone, and reconstructing the Fermi-surface.  These effects are outside of the scope of this paper.  However, some theoretical models of charge-density wave \cite{Wang2014} or pair-density wave \cite{Agterberg2014} also involve TRSB and can produce non-zero PKE from chiral-nematic charge order \cite{Nandkishore2014}.  Possible connection between these models and the loop-current models presented here requires further investigation.

\paragraph{Acknowledgements.}  I am grateful to A.~Kapitulnik and T.~Giamarchi for discussions at the ECRYS-2014 conference; P. Armitage for sharing references to $\rm Cr_2O_3$; P.~Bourges, M.~Greven, and L.~Taillefer for e-mail discussions of the paper; and S. S. Pershoguba for help in preparing figures.

\section*{References}

\end{document}